\documentclass{article}
\usepackage{PRIMEarxiv}

\usepackage[utf8]{inputenc} 
\usepackage[T1]{fontenc}    
\usepackage{hyperref}       
\usepackage{url}            
\usepackage{booktabs}       
\usepackage{amsfonts}       
\usepackage{nicefrac}       
\usepackage{microtype}      
\usepackage{lipsum}
\usepackage{fancyhdr}       
\usepackage{graphicx}       
\usepackage{rotating}       
\usepackage{rotfloat}
\usepackage[acronyms]{glossaries} 

\usepackage{nccmath}
\usepackage{empheq} 
\usepackage{physics} 
\usepackage{color}
\usepackage{subcaption}
\usepackage{braket}
\usepackage{xcolor}
\usepackage{algorithm}
\usepackage{algorithmicx}
\usepackage{algpseudocode}
\usepackage{multicol} 

\newcommand{\dontshow}[1]{} 

\newacronym{cbs}{CBS}{Convergence Born Series}
\newacronym{fft}{FFT}{Fast Fourier Transform}

\makeglossaries
\setkeys{glslink}{hyper=false}  

\title{Automatic skull-template alignment without a guidance image.}

\author{
  Oscar Bates \\
  \textrm{Dept. Earth Science and Engineering} \\
  \textrm{Imperial College London} \\
  \texttt{o.bates18@imperial.ac.uk} \\
  \And
  Carlos Cueto \\
  \textrm{Dept. Earth Science and Engineering} \\
  \textrm{Imperial College London} \\
  \And
  Ciaran Coleman \\
  \textrm{Sonalis Imaging Limited} \\
  \And
  Cameron A. B. Smith \\
  \textrm{Dept. Bioengineering} \\
  \textrm{Imperial College London} \\
  \And
  Lluis Guasch \\
  \textrm{Dept. Earth Science and Engineering} \\
  \textrm{Imperial College London} \\
  \And
  Oscar Calderon Agudo \\
  \textrm{Dept. Earth Science and Engineering} \\
  \textrm{Imperial College London} \\
}

\begin{document}
\maketitle

\begin{abstract}
    Transcranial ultrasound must overcome the significant challenge of the human skull, limiting both imaging and therapeutic applications. While high-fidelity numerical simulations can compensate for skull-induced distortions, they require precise skull templates (typically derived from Computed Tomography, CT) and spatial alignment to the patient’s anatomy. Current alignment relies on concurrent Magnetic Resonance Imaging (MRI) for registration, introducing financial, logistical, and throughput barriers. To overcome these challenges, we present Manifold Optimisation for Full-Waveform Inversion (MOFI), a method to register skull templates without using a guidance image. Our method aligns the skull template by minimising the difference between simulated and observed radio-frequency acoustic data. We demonstrate that MOFI accurately recovers the position of skull templates in silico and in vitro, offering an alternative to MRI guidance-based registration. These results indicate that MOFI has the potential to be a practical alternative to MRI-guided approaches, reducing the barriers to wider clinical adoption of transcranial ultrasound.
\end{abstract}


\begin{multicols}{2}

Transcranial ultrasound encompasses numerous therapeutic and imaging applications. Currently, the major neuroimaging modalities, Magnetic Resonance Imaging (MRI) and X-ray Computed Tomography (CT), are large, expensive, immobile, high-power instruments that are deployable only in specialized facilities. Pulse-Echo Ultrasound is a safe, fast, and portable but transcranial B-mode imaging struggles with limited resolution \cite{estrada_imaging_2025}. Therapeutic ultrasound has few direct analogues in clinical use and encompasses neuromodulation \cite{tufail_transcranial_2010}, drug delivery \cite{hynynen_noninvasive_2001},  and ablation \cite{elias_pilot_2013}.  

Ultrasound neurotechnologies are hindered by the high impedance and heterogeneous structure of the skull, which abberates, scatters, and attenuates the ultrasound wavefield \cite{pichardo_multi-frequency_2010}. These effects substantially degrade both therapeutic targeting \cite{robertson_sensitivity_2017} and transcranial imaging \cite{guasch_full-waveform_2020}.  

Recent work combining wide-aperture transcranial Ultrasound Tomography (t-UST) systems with Full-Waveform Inversion (FWI) has shown promise for recovering intracranial structure \cite{guasch_full-waveform_2020, robins_dual-probe_2023, mitcham_transcranial_2025}. FWI formulates image reconstruction as a non-linear optimisation problem, estimating the acoustic parameter field that best matches measured radio-frequency acoustic data. However, the large impedance contrast between bone and soft tissue continues to pose severe challenges for conventional gradient-based optimisation. FWI is high dimensional, often involving tens to hundreds of millions of elements and is prone convergence issues---such as cycle skipping---that impair the reconstruction \cite{virieux_overview_2009}. 

Various strategies have been proposed to address these limitations, including alternative loss functions \cite{van_leeuwen_mitigating_2013, warner_adaptive_2016} and preconditioning  the image vector via neural networks \cite{orozco_invertiblenetworksjl_2024} or finite-element grids \cite{marty_shape_2023}, though these have yet to be validated on experimental UST data. In therapeutic  ultrasound, a long-standing solution is to estimate the skull's acoustic properties from patient-specific CT data and numerically model skull-induced aberration \cite{aubry_experimental_2003}, enabling corrections to be applied. One analogous approach in t-UST is to use prior information from the skull is provided to simplify the optimisation problem \cite{guasch_full-waveform_2020, robins_dual-probe_2023}. 

However, an important practical challenge is the alignment of skull templates to patient anatomy. Existing approaches typically rely on MRI guidance images to register CT-derived skull models \cite{mcdannold_transcranial_2010}, which undermines the clinical utility of t-UST and complicates deployment. Guidance-free registration remains largely unexplored, with only one prior study using reflected ultrasound time-of-flight data and constrained optimisation, restricting rotations and translations to ±3° and ±3 mm \cite{oreilly_registration_2016}. 

We present Manifold Optimisation for Full-Waveform Inversion (MOFI), a framework for guidance-free skull-template alignment utilising both reflected and transmitted acoustic data employing unconstrained optimisation. MOFI identifies a smooth, low-dimensional manifold that embeds key properties of the FWI image vector, enabling optimisation over physically meaningful transformations while preserving high-fidelity wave propagation modelling. By operating in this reduced space, MOFI effectively preconditions the FWI gradient and improves robustness to large initial misalignments. 

\section{Method}

In standard FWI, the gradient is computed for each pixel of the discretised sound speed grid using an adjoint state method. The geometry of this gradient is inherently tied to the discretisation of the sound speed grid, where the dimensionality of the update is determined by the number of pixels in the image. However, this geometry is not necessarily optimal for FWI updates, as the sound speed discretisation is primarily chosen to ensure the stability of numerical simulations rather than to reflect the geometric properties of the underlying object. We hypothesise that a more geometrically informed choice of update parameters can significantly improve the update direction, enabling access to solutions that are unattainable through standard FWI. This principle underpins MOFI, which leverages a low-dimensional, geometrically meaningful parameterisation to enhance the efficiency and effectiveness of the inversion process.

\subsection{Full-Waveform Inversion}

We consider a standard time-domain acoustic FWI setup. The unknown is the sound-speed distribution \(\mathbf{c}(\mathbf{x})\), \(\mathbf{x} \in \Omega \subset \mathbb{R}^n\). Given appropriate initial and boundary conditions, including a source \(s(\mathbf{x}, t)\), the acoustic wavefield \(\mathbf{u}_p(\mathbf{x}, t)\) satisfies
\[
\Box(\mathbf{c}) \mathbf{u}_p(\mathbf{x}, t) = s(\mathbf{x}, t),
\]
where
\[
\Box(\mathbf{c}) = \frac{1}{\mathbf{c}(\mathbf{x})^2} \frac{\partial^2}{\partial t^2} - \nabla^2,
\]

For a point source at position \(\mathbf{x}_s\) with time signal \(\varsigma(t)\), we write
\[
s(\mathbf{x}, t) = \delta(\mathbf{x} - \mathbf{x}_s)\,\varsigma(t).
\]
The predicted data are obtained by sampling the wavefield at receiver positions \(\mathbf{x}_r\):
\[
\mathbf{d}_p(\mathbf{x}_r, t; \mathbf{c}, \mathbf{x}_s, \varsigma) = \delta(\mathbf{x} - \mathbf{x}_r) \, \mathbf{u}_p(\mathbf{x}, t; \mathbf{c}, \mathbf{x}_s, \varsigma).
\]
Let \(L(\mathbf{c})\) denote the collection of all predicted data for a given model \(\mathbf{c}\) (stacking all receivers, times, and sources into a single data vector), and let \(\mathbf{d}\) be the corresponding observed data. We use the usual least-squares data misfit
\[
f(\mathbf{c}) = \frac{1}{2} \| \mathbf{d} - L(\mathbf{c}) \|_2^2.
\]

The goal of FWI is to find a model \(\mathbf{c}\) that minimises \(f(\mathbf{c})\). This is typically done by iterative gradient-based optimisation. The gradient of \(f\) with respect to \(\mathbf{c}\) is obtained using the adjoint-state method. The gradient is given by:
\[
\frac{\partial f}{\partial \mathbf{c}} = - \mathbf{u}_p(\mathbf{c})^T \, \frac{\partial \Box(\mathbf{c})}{\partial \mathbf{c}} \, \mathbf{u}_{\text{adj}}(\mathbf{c}),
\]
where the adjoint wavefield \(\mathbf{u}_{\text{adj}}(\mathbf{c})\) is the solution of the adjoint problem:
\[
\Box(\mathbf{c})^T \mathbf{u}_{\text{adj}}(\mathbf{c}) = L(\mathbf{c}) - \mathbf{d}.
\]

In practice, each iteration consists of:

\begin{enumerate}
    \item \textbf{Forward solve:} Compute \(\mathbf{u}_p\) and the predicted data \(L(\mathbf{c})\).
    \item \textbf{Adjoint solve:} Back-propagate the data residual \(L(\mathbf{c}) - \mathbf{d}\) to obtain an adjoint wavefield \(\mathbf{u}_{\text{adj}}\).
    \item \textbf{Gradient computation:} Form the zero-lag of the cross-correlation between the forward and adjoint wavefields to obtain the gradient vector \(\partial f / \partial \mathbf{c}\) at each grid point.
\end{enumerate}

The FWI gradient indicates how the sound-speed model should be updated to reduce the misfit. In conventional FWI, we use this gradient directly to perform a pixel-wise update of \(\mathbf{c}\).

\subsection{Manifold Optimisation for FWI} \label{seg}

The standard pixel-wise parameterisation in FWI is inefficient when the mismatch between predicted and observed data arises from geometric deformations, such as global translations or rotations. To address this, MOFI introduces a low-dimensional parameterisation that captures the geometry of the deformation. In this work, we focus on representing the model update through rigid-body motion—namely, translations and rotations of the image — though MOFI can also be extended to non-rigid deformations and other parametrisations. By constraining updates to a geometrically meaningful manifold, MOFI improves the efficiency, accuracy and robustness of the inversion process.

We start from a reference sound-speed image \(\mathbf{c}(\mathbf{x}_{\text{orig}})\) defined on the regular grid \(\mathbf{x}_{\text{orig}}\). Rather than updating \(\mathbf{c}\) directly, we introduce a small set of motion parameters:
\[
\boldsymbol{\phi} = \{\theta, \delta_1, \delta_2\},
\]

which describe a rotation by angle \(\theta\) around a chosen centre and a translation by distance \((\delta_1, \delta_2)\).

These parameters define a rigid transformation of coordinates using the Special Euclidean Matrix Group \(SE(2)\):
\begin{align*}
\mathbf{x}_{\text{orig}}' &= T_{\boldsymbol{\phi}}(\mathbf{x}_{\text{orig}}) \\
\begin{pmatrix}
x_{\text{orig},1}' \\
x_{\text{orig},2}' \\
1
\end{pmatrix}
&=
\begin{pmatrix}
\cos(\theta) & -\sin(\theta) & \delta_1 \\
\sin(\theta) & \cos(\theta) & \delta_2 \\
0 & 0 & 1
\end{pmatrix}
\begin{pmatrix}
x_{\text{orig},1} \\
x_{\text{orig},2} \\
1
\end{pmatrix}.
\end{align*}

Here, \(T_{\boldsymbol{\phi}}\) is the forward transformation matrix that maps the original coordinates \(\mathbf{x}_{\text{orig}}\) to the transformed coordinates \(\mathbf{x}_{\text{orig}}'\).

The aim is to find pixel values for the transformed image. The transformed coordinates \(\mathbf{x}_{\text{orig}}'\) are not usually integers, so \(\mathbf{c}_{\boldsymbol{\phi}}(\mathbf{x}_{\text{orig}}')\) is not defined on a regular pixelated grid. Instead, a pixelated coordinate grid \(\mathbf{x}_{\text{trans}}\) is defined for a ``new'' image \(\mathbf{c}(\mathbf{x}_{\text{trans}})\). Then, each pixel of the new image is associated with a location \(\mathbf{x}_{\text{trans}}'\) on the original image
\[
\mathbf{x}_{\text{trans}}' = T_{\boldsymbol{\phi}}^{-1}(\mathbf{x}_{\text{trans}}).
\]
where the inverse transformation \(T_{\boldsymbol{\phi}}^{-1}\) maps the new coordinate system back to the original coordinate system. The transformed sound-speed image is then parameterised in terms of the original image, given by:
\[
\mathbf{c}_{\boldsymbol{\phi}}(\mathbf{x}_{\text{trans}}') = \mathbf{c}\bigl(T_{\boldsymbol{\phi}}^{-1}(\mathbf{x}_{\text{trans}})\bigr).
\]

Intuitively, \(\mathbf{c}_{\boldsymbol{\phi}}(\mathbf{x}_{\text{trans}}')\) represents the original sound-speed image \(\mathbf{c}(\mathbf{x}_{\text{orig}})\) after it has been rotated by an angle \(\theta\) and translated by \((\delta_1, \delta_2)\). Since the coordinates \(\mathbf{x}_{\text{trans}}' = T_{\boldsymbol{\phi}}^{-1}(\mathbf{x}_{\text{trans}})\) typically do not align with the original grid points, an interpolation algorithm is used to estimate the sound-speed values in the transformed image.

By parametrising the sound-speed image as a function of \(\boldsymbol{\phi}\), we shift the optimisation process from updating individual pixels of \(\mathbf{c}\) to adjusting the geometric transformation parameters. Consequently, the misfit function \(f\) depends on \(\boldsymbol{\phi}\) solely through the transformed model
\[
f(\boldsymbol{\phi}) = f(\mathbf{c}_{\boldsymbol{\phi}}).
\]
enabling a more efficient search direction and geometrically meaningful optimisation process.

Using the chain rule, the gradient of \(f\) with respect to the motion parameters is:
\[
\frac{\partial f}{\partial \boldsymbol{\phi}} = \left( \frac{\partial \mathbf{c}_{\boldsymbol{\phi}}}{\partial \boldsymbol{\phi}} \right)^T \frac{\partial f}{\partial \mathbf{c}_{\boldsymbol{\phi}}}.
\]
Here, \(\partial f / \partial \mathbf{c}_{\boldsymbol{\phi}}\) is the usual FWI gradient vector with respect to the sound speed (computed by the adjoint-state method), and \(\partial \mathbf{c}_{\boldsymbol{\phi}} / \partial \boldsymbol{\phi}\) is a Jacobian that describes how a change in the motion parameters \(\boldsymbol{\phi}\) changes the sound speed \(\mathbf{c}\) at each pixel.

This formula shows that we can obtain the gradient with respect to the motion parameters \(\boldsymbol{\phi}\) by re-weighting and combining the values of the standard FWI gradient.

\subsection{Numerical Implementation} \label{implementation}

The implementation is analogous to rigid registration. Each pixel in the sound-speed image is assigned a coordinate, which is a 2D vector representing its position relative to a chosen centre of rotation. Given the current parameters \(\boldsymbol{\phi} = \{\theta, \delta_1, \delta_2\}\), we construct the corresponding rigid transform \(T_{\boldsymbol{\phi}}\):

\begin{enumerate}
    \item The transformed sound-speed field \(\mathbf{c}_{\boldsymbol{\phi}}(\mathbf{x}_{\text{trans}}')\) is computed by mapping each grid point \(\mathbf{x}_{\text{trans}}\) back to the original domain via \(\mathbf{x}_{\text{trans}}' = T_{\boldsymbol{\phi}}^{-1}(\mathbf{x}_{\text{trans}})\).
    \item Bilinear interpolation is used to evaluate \(\mathbf{c}\) at coordinates \(\mathbf{x}_{\text{trans}}'\) to form \(\mathbf{c}_{\boldsymbol{\phi}}\).
    \item Solve the forward problem to obtain \(L(\mathbf{c}_{\boldsymbol{\phi}})\).
    \item Compute the data residual \(L(\mathbf{c}_{\boldsymbol{\phi}}) - \mathbf{d}\).
    \item Solve the adjoint problem to form the standard FWI gradient vector \(\partial f / \partial \mathbf{c}_{\boldsymbol{\phi}}\).
    \item Obtain the gradient with respect to the motion parameters by applying \(\frac{\partial f}{\partial \boldsymbol{\phi}} = \left( \frac{\partial \mathbf{c}_{\boldsymbol{\phi}}}{\partial \boldsymbol{\phi}} \right)^T \frac{\partial f}{\partial \mathbf{c}_{\boldsymbol{\phi}}}\).
\end{enumerate}

In practice, we do not provide an explicit expression for the Jacobian \(\partial \mathbf{c}_{\boldsymbol{\phi}} / \partial \boldsymbol{\phi}\). Instead, we use reverse-mode automatic differentiation (Pytorch \cite{paszke_pytorch_2019}) to directly compute the product of this Jacobian transpose with the vector \(\partial f / \partial \mathbf{c}_{\boldsymbol{\phi}}\). The FWI gradient with respect to \(\mathbf{c}\) is obtained using the adjoint-state method implemented in Stride \cite{cueto_stride_2022}.

The resulting gradient \(\partial f / \partial \boldsymbol{\phi}\) gives the steepest descent direction in the space of motion parameters. We perform a gradient-based update of \(\boldsymbol{\phi}\), with gradient normalisation and line search to improve stability near the minimum. This allows MOFI to capture the dominant geometric mismatch (translations and rotations) by using a more efficient and accurate search direction, in contrast to the pixel-wise update. Once this coarse alignment is achieved, the standard FWI gradient can, if desired, be used in the usual way to refine smaller-scale features.

\section{Results} \label{results}

\subsection{In-silico} \label{insilico}

\begin{figure*}
\centering \includegraphics[width=0.9\textwidth, bb=0 0 1650 1520, clip=True]{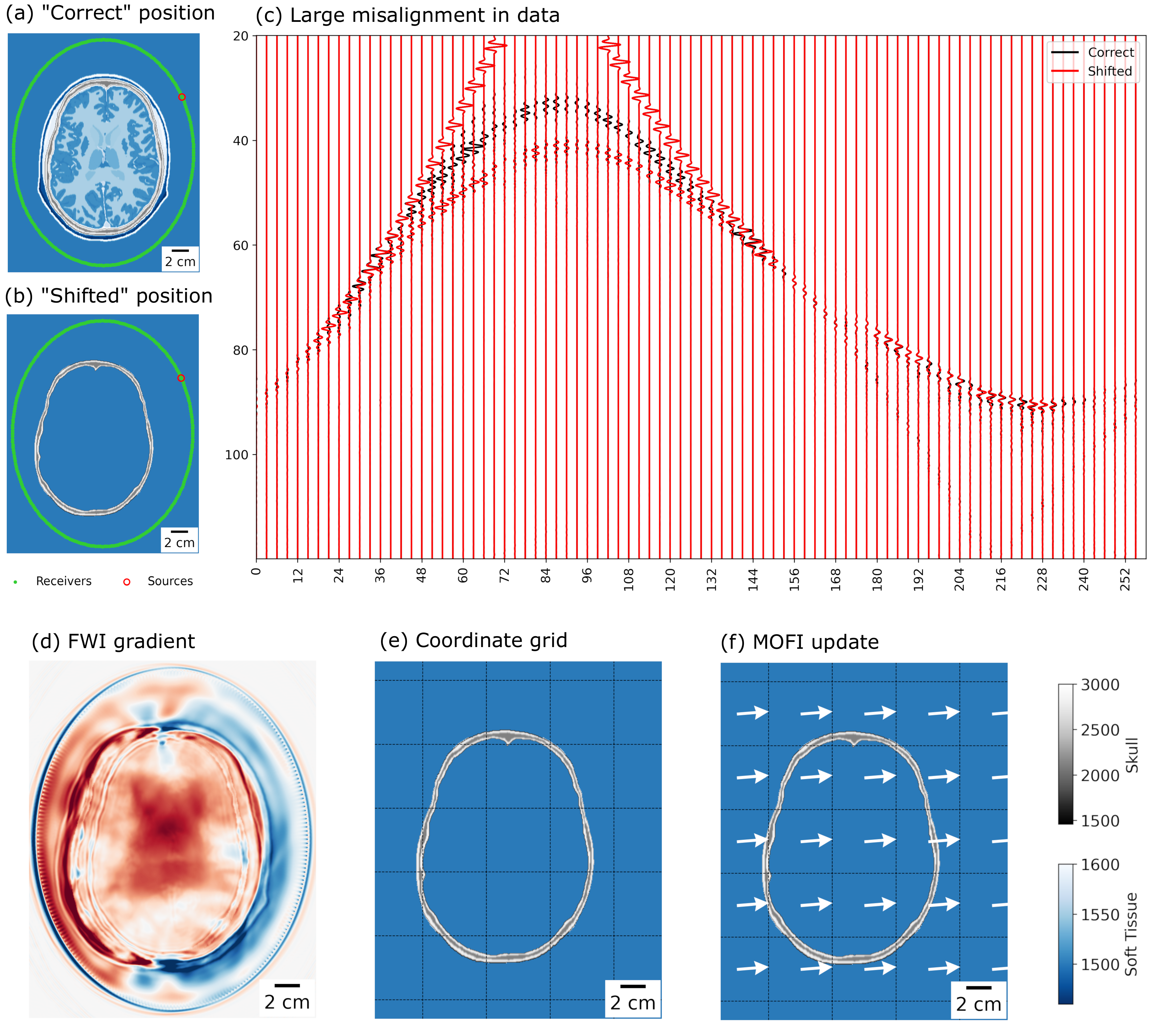} 
\caption{(a) The ground truth model in the ``correct'' position and (b) the skull template in the ``shifted'' position. (c) The correct and shifted data -- respectively the black and red traces -- show a large offset. (d) The FWI gradient captures the leftward shift, but the pixel-wise details are unsatisfactory. (d) By parametrising the image using coordinates, (f) MOFI (Manifold Optimisation for FWI) can represent the update in terms of translation and rotation of the image, providing an update direction (white arrows) that captures the shift.}
\label{fig1}
\end{figure*}

Figure~\ref{fig1}(a) shows an \textit{in-silico} phantom representing the sound speed of the head. This phantom is constructed from the MIDA model \cite{iacono_mida:_2015} and the IT'IS database of tissue properties \cite{hasgall_itis_2015}. The shape of the grid is (560, 450) and isotropic spacing of the grid is $0.5 \, \mathrm{mm}$.

Figure~\ref{fig1} provides an initial validation of our methodology. A large offset between (a) the ``correct'' and (b) the ``shifted'' p-wave velocity models causes (c) a large offset between the corresponding data (black and red traces). Figure~\ref{fig1}(d) shows that the standard FWI gradient is sensitive to this shift, but applying this gradient pixel-wise leads to inefficient updates. MOFI replaces pixel-wise updates with a small set of motion parameters (translations and rotations of the image), which provides a more efficient update direction. Figure~\ref{fig1}(e) is an initial validation of our hypothesis---the white arrows show the direction of the MOFI update, which is the reverse of the shift induced in Figure~\ref{fig1}(b).

\begin{figure*}
\centering \includegraphics[width=0.95\textwidth]{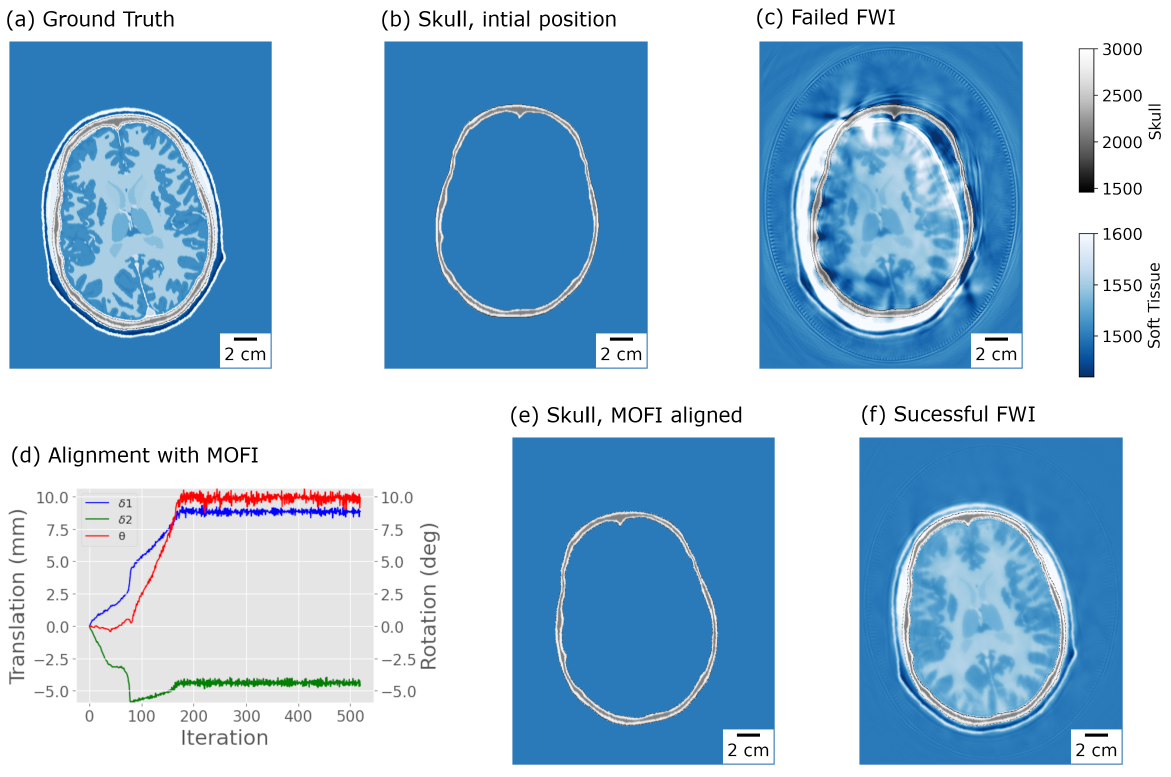}
\caption{In-silico wavefield data are simulated from (a) the ground truth sound speed. (b) The skull template at the initial position produces (c) a failed FWI reconstruction. (d) MOFI identifies the correct translation ($\mathrm{\delta1, \delta2}$) and rotation ($\mathrm{\theta}$). (d) The MOFI aligned skull template produces (f) a successful FWI reconstruction.}
\label{fig2}
\end{figure*}

Figure~\ref{fig2} demonstrates that MOFI is able to accurately locate the skull template when the ground truth sound speed undergoes a substantial shift in position and orientation (Fig~\ref{fig2}(a)). Relative to the the original image, the rotation is $10^\mathrm{o}$---relative to the vertical---about the centre of the image. The translation is $(14, -5.85) \, \mathrm{mm}$.

Figure~\ref{fig2}(b) shows a starting model where the skull template has the correct sound speed and shape, but its position (i.e. translation and rotation) is incorrect. Figure~\ref{fig2}(c) shows that FWI cannot reconstruct the brain from this starting model. Except for the position of the skull, this example is an inverse crime---the modelling and reconstruction physics are identical, the exact source wavelet and transducer positions are known, and the data are not cycle-skipped as the inversion starts from 100kHz.

Figure~\ref{fig2}(d) shows the three parameters \(\boldsymbol{\phi}\) of SE(2) during the optimisation process. The translation starts at $(0,0) \, \mathrm{mm}$ and the rotation starts at $0^\mathrm{o}$. In total there are 500 iterations. The last 200 iterations oscillate around a stationary point, which give a translation $(13.9 \pm 0.3, -6.8 \pm 0.3) \, \mathrm{mm}$ and rotation $9.8 \pm 0.3^\mathrm{o}$. With respect to the known displacement, the horizontal translation is incorrect by 1mm, or two pixels, the other values are accurate to less than a pixel. The  stationary point is reached after approximately 200 iterations, which is computationally feasible. Furthermore, because the group has a limited number of parameters, the problem is---in a sense---overdetermined, which means that a small batch size (10\% of the emitters) can be used. Figure~\ref{fig2}(e) shows the skull template at the optimal position, which is determined by the optimisation on SE(2). The FWI reconstruction in Figure~\ref{fig2}(f) confirms that the starting model is now correctly located.

\subsection{In vitro} \label{invitro} 



\begin{figure*}[htbp]
\centering
\rotatebox{90}{%
  \begin{minipage}{\textheight}
    \centering
    \includegraphics[height=0.7\textheight]{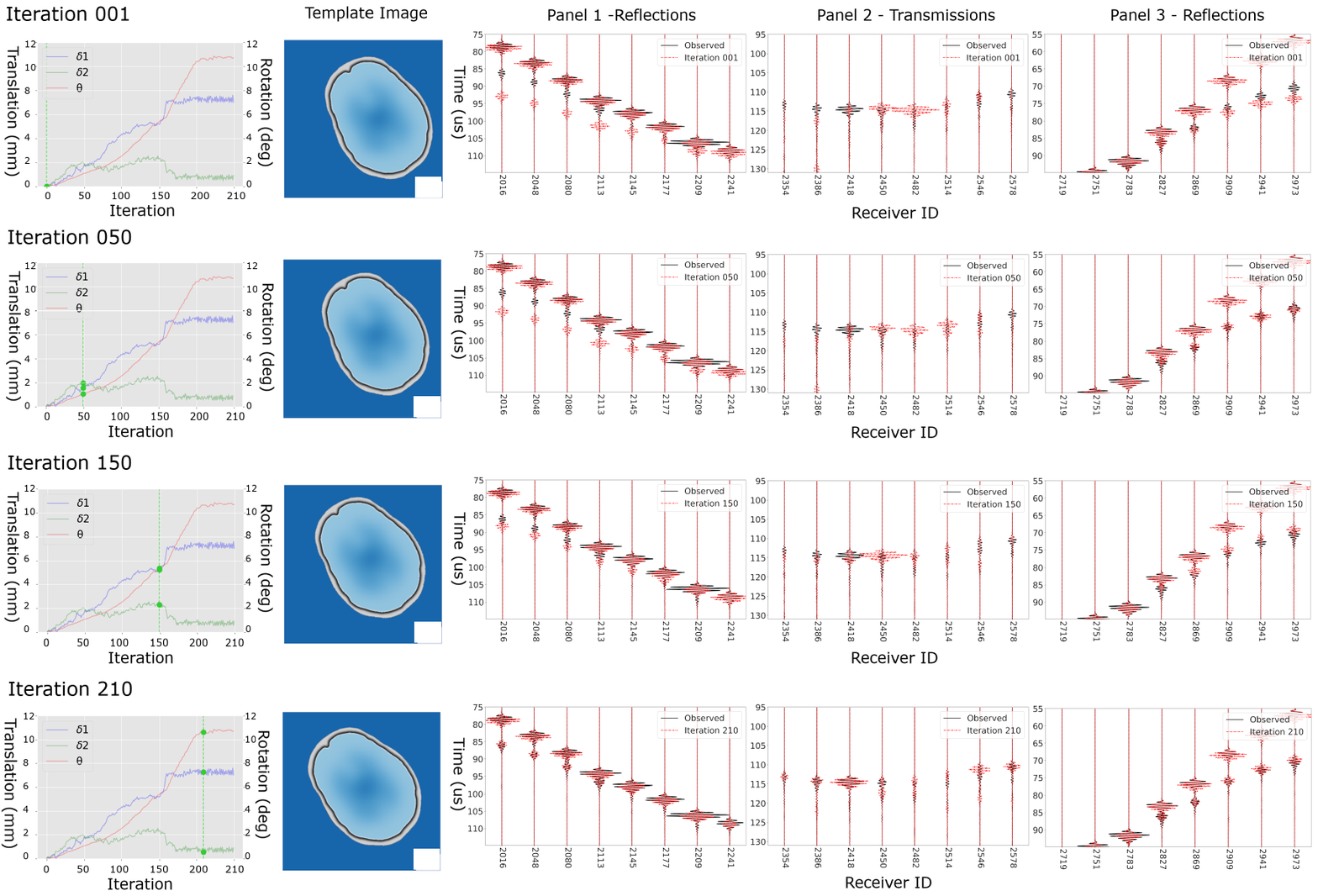}
    \captionof{figure}{MOFI optimisation on the \textit{in vitro} skull template. (column 1) The parameters $(\delta1,\delta2,\theta)$ follow a smooth curve from the initial to the final values, which (column 2) corresponds to a smooth movement from the inital to the final position. (column 3, panel 1) One set of the reflected data are gradually time-shifted to earlier. (column 4, panel 2) The transmitted data are redistributed to adjacent transducers, but not time-shifted. (column 5, panel 3) The second set of reflected data are also time-shifted, but these data overshoot before setting at the correct delay.}
    
    \label{fig3}
  \end{minipage}%
}
\end{figure*}

FWI reconstructions of lab data are substantially more challenging. First, the impulse response of the transducer is unknown. This means that the spatial response of the emitter and sensitivity of the receiver are difficult to determine with precision. Second, the physics of the problem are unknown. To reduce computation time, we solve the acoustic wave equation with constant density and zero attenuation. In reality, materials have variable density, attenuate wave energy, and may support elastic or anisotropic physics. Finally, the properties of the starting model are unknown. Different sources in the literature provide conflicting estimates of the skull sound speed, which suggests that these estimates have wide errors. Furthermore, the shape of the skull can only be determined imprecisely from CT or MRI data (typical resolution of $0.5 \, \mathrm{mm}$), and its position is only known approximately despite careful placement.

Figure~\ref{fig3} demonstrates that MOFI can optimise the position of the \textit{in vitro} skull template. Figure~\ref{fig3}(column 1) shows the parameter values $\boldsymbol{\phi}=(\delta_1, \delta_2, \theta)$ for iterations 1, 50, 150, and 210. These correspond to the sound speed images in Figure~\ref{fig3}(column 2), which show the template moving smoothly from the initial to the final position.


Figures~\ref{fig3}(columns 3-5) provides a detailed view of the data misfit during the optimisation process. The observed acoustic data are in black, the predicted acosutic data are in red.

Broadly, transmitted energy in the predicted data (panel 2, red traces) is smoothly shifted to adjacent receivers, but the arrival time remains unchanged. In contrast, reflected energy (panel 1\&3, red traces, second arrival) is shifted to earlier arrival times. The first arrival in panels 1\&3 travels directly through water and remains unchanged throughout the optimisation. In detail:
\begin{itemize}
    \item Figure~\ref{fig3}, Iteration 1:
    The predicted and observed data exhibit significant misalignment across all panels.
    \item Figure~\ref{fig3}, Iteration 50:
    In panels 1 and 3, the reflected predicted data (second arrivals) shift to earlier times, moving toward the observed data. In panel 1, a complete alignment in phase and amplitude is achieved, while the transmitted data in panel 2 remain largely unchanged.
    \item Figure~\ref{fig3}, Iteration 150:
    In panel 3, the predicted reflected data overshoot the observed data, while in panel 1 (on the opposite side of the array) the arrival time continues to steadily decrease. In panel 2, the timing of the transmitted data remains stable, though energy is redistributed to adjacent receivers.
    \item Figure~\ref{fig3}, Iteration 210:
    The data in all three panels align closely: transmitted energy is correctly redistributed, and reflected energy is appropriately time-shifted. Minor misalignments persist, particularly for reflected arrivals in panel 1, which are cycle-skipped.
\end{itemize}

\begin{figure*}
\centering 
\includegraphics[width=0.99\textwidth]{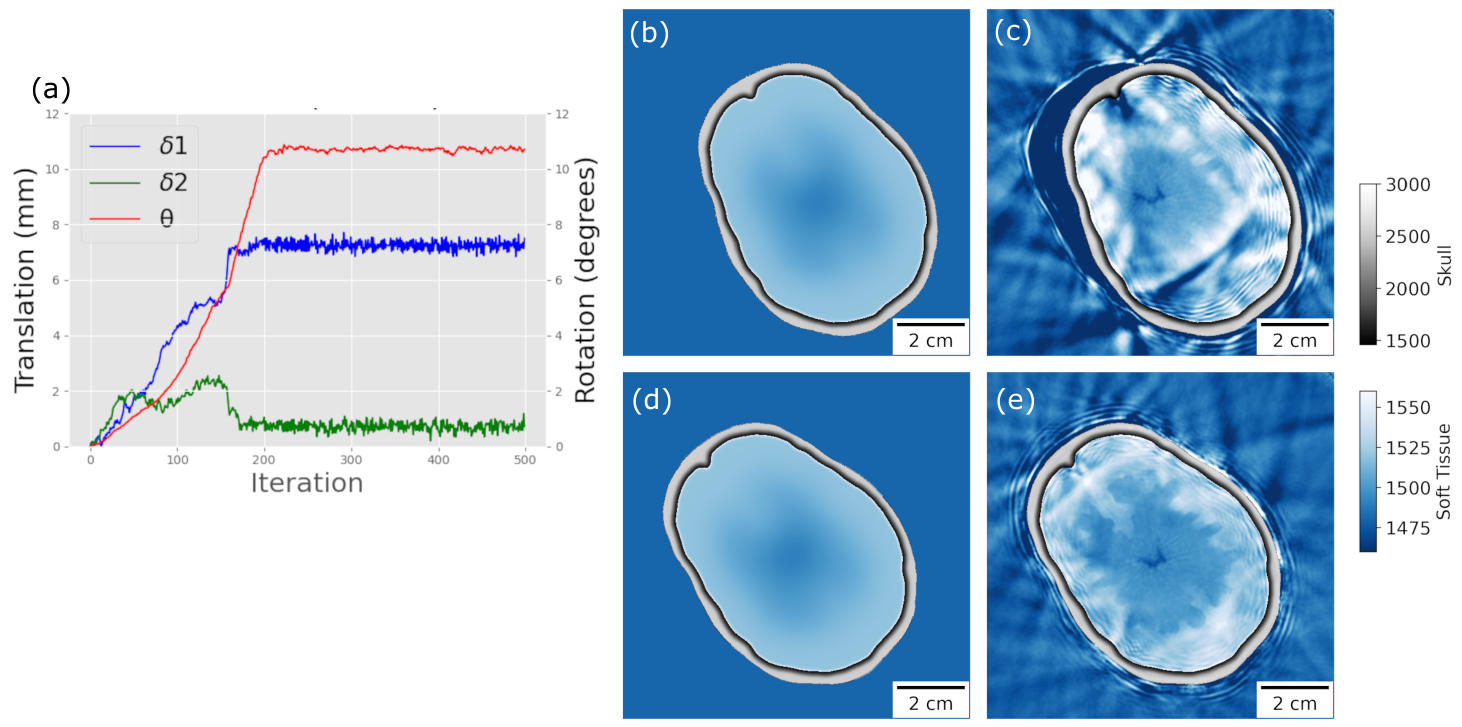}
\caption{(a) MOFI identifies a translation \(\boldsymbol{\delta}=(7.2 \pm 0.2, 0.7 \pm 0.2) \, \mathrm{mm}\) and a rotation $\mathrm{\theta} = 10.71 \pm 0.07^\mathrm{o}$. (b) The skull template at the initial position produces (c) a failed FWI reconstruction. (d) The MOFI aligned skull template provides (e) a successful FWI reconstruction.}
\label{fig5}
\end{figure*}

Figure~\ref{fig5} demonstrates that MOFI itentifies an accurate position and orientation for the skull template. Figure~\ref{fig5}(a) shows that the optimal translation is $(7.2 \pm 0.2, 0.7 \pm 0.2) \, \mathrm{mm}$ and the optimal rotation is $10.71 \pm 0.07^\mathrm{o}$, measured from the final 300 iterations. The stability of the translation and rotation values between iteration 200 and iteration 500 confirms that the optimisation process has converged. Figure~\ref{fig5}(b) shows the skull template in its initial position. Previous work \cite{robins_dual-probe_2023} shows that this template has approximately the correct sound speed and shape, but none of the values are known precisely. Figure~\ref{fig5}(c) confirms the incorrect placement because the FWI reconstruction has failed. Figure~\ref{fig3}(d) shows the skull template in the optimal position, identified at iteration 500. Figure~\ref{fig5}(e) shows an accurate FWI reconstruction, which confirms that the starting model is now correctly positioned.

\section{Discussion}
\label{discussion}

Our approach delivers strong results whenever an approximate sound-speed template is available and requires alignment. Ultrasound neurotechnologies rely on accurate skull template alignment, as misalignment compromises the precision of numerical refocusing corrections. Rigid MOFI addresses this challenge by enabling automatic skull template alignment with sub-pixel accuracy. This innovation reduces the need for manual intervention in FWI neuroimaging and provides a robust alternative — or complement — to MRI-guided image registration in transcranial focused ultrasound therapy.

The effectiveness of MOFI stems from its use of the Special Euclidean Group SE(2) to constrain the full-waveform inversion (FWI) optimisation process. SE(2) is part of a broader class of mathematical structures known as Lie Groups, which combine algebraic group properties with the smooth geometry of manifolds. These groups are particularly suited to modelling continuous transformations---such as rotations and translations---while preserving geometric symmetries like angles and distances. As a result, Lie Groups offer a natural framework for optimisation problems. Their utility is well-documented in fields such as image registration and robotics. For instance, diffeomorphic non-rigid image registration, which leverages the Diffeomorphism group, significantly outperforms free-form deformation methods when handling large deformations \cite{beg_computing_2005}. In robotics, manifold optimisation using Lie Groups enhances the accuracy of trajectory mapping \cite{brockett_robotic_1984, horn_closed-form_1987, sola_micro_2021}.

Another perspective emerges from convex optimisation, where computational complexity scales polynomially with the number of dimensions \cite{boyd_convex_2004}. Although full-waveform inversion (FWI) is non-convex, the convex case implies that the complexity of non-convex optimisation also increases with dimensionality. Conventional FWI optimisation algorithms operate on a discretised sound-speed grid, where the grid is designed to ensure the stability of the numerical simulator rather than optimisation efficiency. In Figures \ref{fig2} and \ref{fig5}, the discretised sound-speed grids consist of 252,000 and 672,000 pixels, respectively — each pixel representing an individual dimension in the optimisation space.

Thus, by constraining the optimisation to just three dimensions—rotation and translation—MOFI circumvents the challenges posed by high-dimensional discretised grids, such as those in Figures 2 and 5, where each pixel introduces an additional optimisation variable. Dimensionality reduction by manifold optimisation is not unique to FWI \cite{kadu_salt_2017, marty_shape_2023} but, through the use of SE(2), the Lie Group structure of MOFI ensures a smooth optimisation manifold that retains algebraic group properties. This contrasts with level set methods \cite{kadu_salt_2017}, which lack a group structure and may struggle to enforce specific deformations. Furthermore, unlike methods that invert for finite-element grid control points \cite{marty_shape_2023}, MOFI guarantees invertibility, preventing grid points from self-intersecting and preserving the image’s topology. As a result, MOFI delivers a robust and efficient optimisation process, readily adaptable to any discretisation (e.g. finite-difference or finite-element ) while maintaining stability for large deformations.


Looking ahead, we aim to explore more expressive Lie Groups, such as the Diffeomorphism Group, and investigate how concepts from Lie Groups could enhance Deep Learning approaches. The design of Deep Learning models often results in non-smooth representations — exemplified by the challenges in training Generative Adversarial Networks. Approaches that regularise inverse problems but do not explicitly enforce smoothness, such as the Deep Image Prior \cite{ulyanov_deep_2018} or Score-based Generative Models \cite{song_solving_2022}, may therefore benefit from incorporating this principle. Similarly, Normalising Flows, which employ smooth, invertible transformations, have already demonstrated their utility in FWI \cite{sun_enabling_2024, orozco_invertiblenetworksjl_2024}.

It is well documented that FWI struggles to converge when the predicted and observed data arrive out-of-phase \cite{virieux_overview_2009}. MOFI is much more resilient than the typical requirement that the phase difference is less than \(90^{\textrm{o}}\), but it will eventually struggle if this distance becomes too large. Therefore, MOFI also benefits from the multi-scale method \cite{bunks_multiscale_1995}. Similarly, the current loss function uses the L2-norm, but alternative formulations from the FWI literature could further expand the basin of attraction. In terms of optimisation parameters, both fixed-step and explicit line-search methods were evaluated, with explicit line-search providing greater stability at the optimum. Gradient normalisation was found to be essential, and using the norm of the first gradient proved most effective, whereas normalising by the maximum of the current gradient led to instability. Future work should also focus on extending this approach to 3D in-vivo datasets to assess its clinical efficacy.

\section{Conclusion}

Our findings demonstrate that a skull sound-speed template can be accurately and automatically located using only acoustic data. This eliminates the need for manual alignment of the ``starting model'' in ultrasound neuroimaging and potentially enhances the image-to-image alignment used during ultrasound therapy. We greatly improve the convergence properties of FWI by leveraging the Special Euclidean Lie Group, which reduces the dimensionality of the FWI optimisation problem while preserving a smooth optimisation manifold. This strategy has been successful in other fields, particularly non-rigid image registration, and suggests that more expressive Lie Groups could entirely eliminate the need for patient-specific prior information in wave-based image-reconstruction algorithms. Together, these findings support the use of MOFI as a principled pathway toward practical, guidance-free transcranial Ultrasound Tomography.

\section{Acknowledgements}
This work was funded by EPSRC grant number EP/X033651/1. Oscar Calderon-Agudo was funded by MRC, Future Leader Fellowship (FLF), MR/V024086/1. We would like to thank the team at Sonalis Imaging Limited for their support and insightful discussions.


\bibliography{MOFIArxiv.bib}{}

@inproceedings{paszke_pytorch_2019,
	title = {{PyTorch}: {An} {Imperative} {Style}, {High}-{Performance} {Deep} {Learning} {Library}},
	volume = {32},
	shorttitle = {{PyTorch}},
	urldate = {2022-08-26},
	booktitle = {Advances in {Neural} {Information} {Processing} {Systems}},
	publisher = {Curran Associates, Inc.},
	author = {Paszke, Adam and Gross, Sam and Massa, Francisco and Lerer, Adam and Bradbury, James and Chanan, Gregory and Killeen, Trevor and Lin, Zeming and Gimelshein, Natalia and Antiga, Luca and Desmaison, Alban and Kopf, Andreas and Yang, Edward and DeVito, Zachary and Raison, Martin and Tejani, Alykhan and Chilamkurthy, Sasank and Steiner, Benoit and Fang, Lu and Bai, Junjie and Chintala, Soumith},
	year = {2019},
}

@article{cueto_stride_2022,
	title = {Stride: {A} flexible software platform for high-performance ultrasound computed tomography},
	volume = {221},
	issn = {0169-2607},
	shorttitle = {Stride},
	doi = {10.1016/j.cmpb.2022.106855},
	language = {en},
	urldate = {2022-08-19},
	journal = {Computer Methods and Programs in Biomedicine},
	author = {Cueto, Carlos and Bates, Oscar and Strong, George and Cudeiro, Javier and Luporini, Fabio and Calderon Agudo, Oscar and Gorman, Gerard and Guasch, Lluis and Tang, Meng-Xing},
	month = jun,
	year = {2022},
	pages = {106855},
}

@book{boyd_convex_2004,
	title = {Convex {Optimization}},
	isbn = {978-0-521-83378-3},
	language = {en},
	publisher = {Cambridge University Press},
	author = {Boyd, Stephen P. and Vandenberghe, Lieven},
	month = mar,
	year = {2004},
	note = {Google-Books-ID: mYm0bLd3fcoC},
}

@inproceedings{ulyanov_deep_2018,
	title = {Deep {Image} {Prior}},
	urldate = {2025-05-30},
	author = {Ulyanov, Dmitry and Vedaldi, Andrea and Lempitsky, Victor},
	year = {2018},
	pages = {9446--9454},
}

@misc{song_solving_2022,
	title = {Solving {Inverse} {Problems} in {Medical} {Imaging} with {Score}-{Based} {Generative} {Models}},
	doi = {10.48550/arXiv.2111.08005},
	urldate = {2025-05-30},
	publisher = {arXiv},
	author = {Song, Yang and Shen, Liyue and Xing, Lei and Ermon, Stefano},
	month = jun,
	year = {2022},
	note = {arXiv:2111.08005 [eess]},
}

@misc{sola_micro_2021,
	title = {A micro {Lie} theory for state estimation in robotics},
	doi = {10.48550/arXiv.1812.01537},
	urldate = {2025-05-30},
	publisher = {arXiv},
	author = {Solà, Joan and Deray, Jeremie and Atchuthan, Dinesh},
	month = dec,
	year = {2021},
	note = {arXiv:1812.01537 [cs]},
}

@article{guasch_full-waveform_2020,
	title = {Full-waveform inversion imaging of the human brain},
	volume = {3},
	copyright = {2020 The Author(s)},
	issn = {2398-6352},
	doi = {10.1038/s41746-020-0240-8},
	language = {en},
	number = {1},
	urldate = {2020-06-03},
	journal = {npj Digital Medicine},
	author = {Guasch, Lluis and Calderon Agudo, Oscar and Tang, Meng-Xing and Nachev, Parashkev and Warner, Michael},
	month = mar,
	year = {2020},
	pages = {1--12},
}

@article{horn_closed-form_1987,
	title = {Closed-form solution of absolute orientation using unit quaternions},
	volume = {4},
	copyright = {© 1987 Optical Society of America},
	issn = {1520-8532},
	doi = {10.1364/JOSAA.4.000629},
	language = {EN},
	number = {4},
	urldate = {2025-04-22},
	journal = {JOSA A},
	publisher = {Optica Publishing Group},
	author = {Horn, Berthold K. P.},
	month = apr,
	year = {1987},
	pages = {629--642},
}

@inproceedings{brockett_robotic_1984,
	address = {Berlin, Heidelberg},
	title = {Robotic manipulators and the product of exponentials formula},
	isbn = {978-3-540-38826-5},
	doi = {10.1007/BFb0031048},
	language = {en},
	booktitle = {Mathematical {Theory} of {Networks} and {Systems}},
	publisher = {Springer},
	author = {Brockett, R. W.},
	editor = {Fuhrmann, P. A.},
	year = {1984},
	pages = {120--129},
}

@article{beg_computing_2005,
	title = {Computing {Large} {Deformation} {Metric} {Mappings} via {Geodesic} {Flows} of {Diffeomorphisms}},
	volume = {61},
	issn = {1573-1405},
	doi = {10.1023/B:VISI.0000043755.93987.aa},
	language = {en},
	number = {2},
	urldate = {2025-04-22},
	journal = {International Journal of Computer Vision},
	author = {Beg, M. Faisal and Miller, Michael I. and Trouvé, Alain and Younes, Laurent},
	month = feb,
	year = {2005},
	pages = {139--157},
}

@article{orozco_invertiblenetworksjl_2024,
	title = {{InvertibleNetworks}.jl: {A} {Julia} package for scalable normalizing flows},
	volume = {9},
	shorttitle = {{InvertibleNetworks}.jl},
	doi = {10.21105/joss.06554},
	urldate = {2025-04-22},
	journal = {The Journal of Open Source Software},
	author = {Orozco, Rafael and Witte, Philipp and Louboutin, Mathias and Siahkoohi, Ali and Rizzuti, Gabrio and Peters, Bas and Herrmann, Felix},
	month = jul,
	year = {2024},
	note = {ADS Bibcode: 2024JOSS....9.6554O},
	pages = {6554},
}

@article{pichardo_multi-frequency_2010,
	title = {Multi-frequency characterization of the speed of sound and attenuation coefficient for longitudinal transmission of freshly excised human skulls},
	volume = {56},
	issn = {0031-9155},
	doi = {10.1088/0031-9155/56/1/014},
	language = {en},
	number = {1},
	urldate = {2025-04-11},
	journal = {Physics in Medicine \& Biology},
	author = {Pichardo, Samuel and Sin, Vivian W and Hynynen, Kullervo},
	month = dec,
	year = {2010},
	pages = {219},
}

@inproceedings{marty_shape_2023,
	title = {Shape optimization for transcranial ultrasound computed tomography},
	volume = {12470},
	doi = {10.1117/12.2654328},
	urldate = {2025-04-11},
	booktitle = {Medical {Imaging} 2023: {Ultrasonic} {Imaging} and {Tomography}},
	publisher = {SPIE},
	author = {Marty, Patrick and Boehm, Christian and Fichtner, Andreas},
	month = apr,
	year = {2023},
	pages = {77--88},
}

@article{sun_enabling_2024,
	title = {Enabling uncertainty quantification in a standard full-waveform inversion method using normalizing flows},
	volume = {89},
	issn = {0016-8033},
	doi = {10.1190/geo2023-0755.1},
	number = {5},
	urldate = {2025-04-11},
	journal = {GEOPHYSICS},
	publisher = {Society of Exploration Geophysicists},
	author = {Sun, Changxiao and Malcolm, Alison and Kumar, Rajiv and Mao, Weijian},
	month = sep,
	year = {2024},
	pages = {R493--R507},
}

@article{van_leeuwen_mitigating_2013,
	title = {Mitigating local minima in full-waveform inversion by expanding the search space},
	volume = {195},
	issn = {0956-540X},
	doi = {10.1093/gji/ggt258},
	number = {1},
	urldate = {2025-04-11},
	journal = {Geophysical Journal International},
	author = {van Leeuwen, Tristan and Herrmann, Felix J.},
	month = oct,
	year = {2013},
	pages = {661--667},
}

@article{warner_adaptive_2016,
	title = {Adaptive waveform inversion: {Theory}},
	volume = {81},
	issn = {0016-8033},
	shorttitle = {Adaptive waveform inversion},
	doi = {10.1190/geo2015-0387.1},
	number = {6},
	urldate = {2025-04-11},
	journal = {GEOPHYSICS},
	publisher = {Society of Exploration Geophysicists},
	author = {Warner, Michael and Guasch, Lluís},
	month = nov,
	year = {2016},
	pages = {R429--R445},
}

@article{aubry_experimental_2003,
	title = {Experimental demonstration of noninvasive transskull adaptive focusing based on prior computed tomography scans},
	volume = {113},
	issn = {0001-4966},
	doi = {10.1121/1.1529663},
	number = {1},
	urldate = {2025-04-11},
	journal = {The Journal of the Acoustical Society of America},
	author = {Aubry, J.-F. and Tanter, M. and Pernot, M. and Thomas, J.-L. and Fink, M.},
	month = jan,
	year = {2003},
	pages = {84--93},
}

@article{robins_dual-probe_2023,
	title = {Dual-{Probe} {Transcranial} {Full}-{Waveform} {Inversion}: {A} {Brain} {Phantom} {Feasibility} {Study}},
	volume = {49},
	issn = {0301-5629},
	shorttitle = {Dual-{Probe} {Transcranial} {Full}-{Waveform} {Inversion}},
	doi = {10.1016/j.ultrasmedbio.2023.06.001},
	number = {10},
	urldate = {2025-04-11},
	journal = {Ultrasound in Medicine \& Biology},
	author = {Robins, Thomas Caradoc and Cueto, Carlos and Cudeiro, Javier and Bates, Oscar and Agudo, Oscar Calderon and Strong, George and Guasch, Lluis and Warner, Michael and Tang, Meng-Xing},
	month = oct,
	year = {2023},
	pages = {2302--2315},
}

@misc{hasgall_itis_2015,
	title = {{IT}’{IS} {Database} for thermal and electromagnetic parameters of biological tissues},
	doi = {10.13099/VIP21000-04-0},
	language = {en},
	urldate = {2019-01-15},
	publisher = {IT'IS Foundation},
	author = {Hasgall, PA and Neufeld, E and Gosselin, MC and Klingenbock, A and Kuster, N},
	year = {2015},
}

@article{virieux_overview_2009,
	title = {An overview of full-waveform inversion in exploration geophysics},
	volume = {74},
	issn = {0016-8033},
	doi = {10.1190/1.3238367},
	number = {6},
	urldate = {2019-02-02},
	journal = {GEOPHYSICS},
	author = {Virieux, J. and Operto, S.},
	month = nov,
	year = {2009},
	pages = {WCC1--WCC26},
}

@article{iacono_mida:_2015,
	title = {{MIDA}: {A} {Multimodal} {Imaging}-{Based} {Detailed} {Anatomical} {Model} of the {Human} {Head} and {Neck}},
	volume = {10},
	issn = {1932-6203},
	shorttitle = {{MIDA}},
	doi = {10.1371/journal.pone.0124126},
	language = {en},
	number = {4},
	urldate = {2019-01-20},
	journal = {PLOS ONE},
	author = {Iacono, Maria Ida and Neufeld, Esra and Akinnagbe, Esther and Bower, Kelsey and Wolf, Johanna and Oikonomidis, Ioannis Vogiatzis and Sharma, Deepika and Lloyd, Bryn and Wilm, Bertram J. and Wyss, Michael and Pruessmann, Klaas P. and Jakab, Andras and Makris, Nikos and Cohen, Ethan D. and Kuster, Niels and Kainz, Wolfgang and Angelone, Leonardo M.},
	month = apr,
	year = {2015},
	pages = {e0124126},
}

@article{bunks_multiscale_1995,
	title = {Multiscale seismic waveform inversion},
	volume = {60},
	issn = {0016-8033},
	doi = {10.1190/1.1443880},
	language = {en},
	number = {5},
	urldate = {2020-06-03},
	journal = {Geophysics},
	author = {Bunks, Carey and Saleck, Fatimetou M. and Zaleski, S. and Chavent, G.},
	month = oct,
	year = {1995},
	pages = {1457--1473},
}

@article{oreilly_registration_2016,
	title = {Registration of human skull computed tomography data to an ultrasound treatment space using a sparse high frequency ultrasound hemispherical array},
	volume = {43},
	copyright = {© 2016 American Association of Physicists in Medicine},
	issn = {2473-4209},
	doi = {10.1118/1.4960362},
	language = {en},
	number = {9},
	urldate = {2026-01-08},
	journal = {Medical Physics},
	author = {O'Reilly, Meaghan A. and Jones, Ryan M. and Birman, Gabriel and Hynynen, Kullervo},
	year = {2016},
	note = {\_eprint: https://aapm.onlinelibrary.wiley.com/doi/pdf/10.1118/1.4960362},
	pages = {5063--5071},
}

@article{mcdannold_transcranial_2010,
	title = {Transcranial {Magnetic} {Resonance} {Imaging}– {Guided} {Focused} {Ultrasound} {Surgery} of {Brain} {Tumors}: {Initial} {Findings} in 3 {Patients}},
	volume = {66},
	issn = {0148-396X},
	shorttitle = {Transcranial {Magnetic} {Resonance} {Imaging}– {Guided} {Focused} {Ultrasound} {Surgery} of {Brain} {Tumors}},
	doi = {10.1227/01.NEU.0000360379.95800.2F},
	language = {en-US},
	number = {2},
	urldate = {2026-01-08},
	journal = {Neurosurgery},
	author = {McDannold, Nathan and Clement, Greg T. and Black, Peter and Jolesz, Ferenc and Hynynen, Kullervo},
	month = feb,
	year = {2010},
	pages = {323},
}

@article{mitcham_transcranial_2025,
	title = {Transcranial ultrasound tomography for brain imaging: {Ex} vivo results and potential for stroke imaging},
	volume = {52},
	issn = {0094-2405, 2473-4209},
	shorttitle = {Transcranial ultrasound tomography for brain imaging},
	doi = {10.1002/mp.18090},
	language = {en},
	number = {10},
	urldate = {2026-01-08},
	journal = {Medical Physics},
	author = {Mitcham, Trevor and Ali, Rehman and Singh, Melanie and McConnell, Sarah and Rahmani, Redi and Schartz, Derrek and Bender, Matthew and Vates, Edward and Duric, Neb},
	month = oct,
	year = {2025},
	pages = {e18090},
}

@article{robertson_sensitivity_2017,
	title = {Sensitivity of simulated transcranial ultrasound fields to acoustic medium property maps},
	volume = {62},
	issn = {0031-9155},
	doi = {10.1088/1361-6560/aa5e98},
	language = {en},
	number = {7},
	urldate = {2026-01-08},
	journal = {Physics in Medicine \& Biology},
	publisher = {IOP Publishing},
	author = {Robertson, James and Martin, Eleanor and Cox, Ben and Treeby, Bradley E},
	month = mar,
	year = {2017},
	pages = {2559},
}

@article{kadu_salt_2017,
	title = {Salt {Reconstruction} in {Full}-{Waveform} {Inversion} {With} a {Parametric} {Level}-{Set} {Method}},
	volume = {3},
	issn = {2333-9403},
	doi = {10.1109/TCI.2016.2640761},
	number = {2},
	urldate = {2026-01-15},
	journal = {IEEE Transactions on Computational Imaging},
	author = {Kadu, Ajinkya and van Leeuwen, Tristan and Mulder, Wim A.},
	month = jun,
	year = {2017},
	pages = {305--315},
}

@article{estrada_imaging_2025,
	title = {Imaging the brain by traversing the skull with light and sound},
	volume = {9},
	copyright = {2025 Springer Nature Limited},
	issn = {2157-846X},
	doi = {10.1038/s41551-025-01433-5},
	language = {en},
	number = {10},
	urldate = {2026-01-19},
	journal = {Nature Biomedical Engineering},
	publisher = {Nature Publishing Group},
	author = {Estrada, Héctor and Deffieux, Thomas and Robin, Justine and Tanter, Mickaël and Razansky, Daniel},
	month = oct,
	year = {2025},
	pages = {1574--1590},
}

@article{hynynen_noninvasive_2001,
	title = {Noninvasive {MR} {Imaging}–guided {Focal} {Opening} of the {Blood}-{Brain} {Barrier} in {Rabbits}},
	volume = {220},
	issn = {0033-8419},
	doi = {10.1148/radiol.2202001804},
	number = {3},
	urldate = {2026-01-19},
	journal = {Radiology},
	publisher = {Radiological Society of North America},
	author = {Hynynen, Kullervo and McDannold, Nathan and Vykhodtseva, Natalia and Jolesz, Ferenc A.},
	month = sep,
	year = {2001},
	pages = {640--646},
}

@article{elias_pilot_2013,
	title = {A {Pilot} {Study} of {Focused} {Ultrasound} {Thalamotomy} for {Essential} {Tremor}},
	volume = {369},
	issn = {0028-4793},
	doi = {10.1056/NEJMoa1300962},
	number = {7},
	urldate = {2026-01-19},
	journal = {New England Journal of Medicine},
	publisher = {Massachusetts Medical Society},
	author = {Elias, W. Jeffrey and Huss, Diane and Voss, Tiffini and Loomba, Johanna and Khaled, Mohamad and Zadicario, Eyal and Frysinger, Robert C. and Sperling, Scott A. and Wylie, Scott and Monteith, Stephen J. and Druzgal, Jason and Shah, Binit B. and Harrison, Madaline and Wintermark, Max},
	month = aug,
	year = {2013},
	note = {\_eprint: https://www.nejm.org/doi/pdf/10.1056/NEJMoa1300962},
	pages = {640--648},
}

@article{tufail_transcranial_2010,
	title = {Transcranial {Pulsed} {Ultrasound} {Stimulates} {Intact} {Brain} {Circuits}},
	volume = {66},
	issn = {0896-6273},
	doi = {10.1016/j.neuron.2010.05.008},
	language = {English},
	number = {5},
	urldate = {2026-01-19},
	journal = {Neuron},
	publisher = {Elsevier},
	author = {Tufail, Yusuf and Matyushov, Alexei and Baldwin, Nathan and Tauchmann, Monica L. and Georges, Joseph and Yoshihiro, Anna and Tillery, Stephen I. Helms and Tyler, William J.},
	month = jun,
	year = {2010},
	pages = {681--694},
}
\bibliographystyle{ieeetr}

\end{multicols}

\end{document}